\documentclass[fleqn,usenatbib]{mnras}

\usepackage{newtxtext,newtxmath}

\usepackage[T1]{fontenc}

\usepackage{graphicx,times}             
\usepackage{natbib}
\usepackage{amsmath}
\bibpunct{(}{)}{;}{a}{}{,}

\usepackage{graphicx,times}
\usepackage{natbib}
\usepackage{adjustbox}
\usepackage{siunitx}
\usepackage{threeparttable}

\usepackage{booktabs}
\usepackage{caption}

\usepackage{amsmath} 

\usepackage{float}

\usepackage{siunitx}
\usepackage{orcidlink}
\usepackage{soul, color, xcolor}
\soulregister{\cite}7 
\soulregister{\citep}7 
\soulregister{\citet}7 
\soulregister{\ref}7 
\soulregister{\pageref}7

\DeclareRobustCommand{\VAN}[3]{#2}
\let\VANthebibliography\thebibliography
\def\thebibliography{\DeclareRobustCommand{\VAN}[3]{##3}\VANthebibliography}

\usepackage{graphicx}	
\usepackage{amsmath}	

\title[Energy-resolved polarisation study of the Crab Nebula with IXPE]{Energy-resolved polarisation study of the Crab Nebula with IXPE}

\author[Wenhao Wei et al. ]{
Wenhao Wei $^{\orcidlink{0009-0000-2414-9449}}$,$^{1}$
Fei Xie $^{\orcidlink{0000-0002-0105-5826}}$,$^{1,2}$ \thanks{E-mail: xief@gxu.edu.cn}
Fabio La Monaca  $^{\orcidlink{0000-0001-8916-4156}}$,$^{2,3}$
Wei Deng $^{\orcidlink{0000-0002-9370-4079}}$,$^{1}$
Mingyu Ge $^{\orcidlink{0000-0002-3776-4536}}$,$^{4,5}$
Kuan Liu $^{\orcidlink{0009-0007-8686-9012}}$,$^{1}$ \newauthor
Chao Zuo $^{\orcidlink{0009-0007-5244-2379}}$,$^{1}$ 
and Wei Chen $^{\orcidlink{0000-0002-5965-7432}}$ $^{1}$
\\
$^{1}$Guangxi Key Laboratory for Relativistic Astrophysics, School of Physical Science and Technology, Guangxi University, Nanning 530004, China; \\
$^{2}$INAF Istituto di Astrofisica e Planetologia Spaziali, Via del Fosso del Cavaliere 100, 00133 Roma, Italy\\
$^{3}$Dipartimento di Fisica, Università degli Studi di Roma “Tor Vergata,” Via della Ricerca Scientifica 1, 00133 Roma, Italy\\
$^{4}$ Key Laboratory of Particle Astrophysics, Institute of High Energy Physics, Chinese Academy of Sciences, Beijing 100049, China\\
$^{5}$ University of Chinese Academy of Sciences, Chinese Academy of Sciences, Beijing 100049, China 
}

\date{Accepted XXX. Received YYY; in original form ZZZ}

\pubyear{\the\year{}}

\usepackage[left]{lineno}
\begin{document}
\label{firstpage}
\pagerange{\pageref{firstpage}--\pageref{lastpage}}
\maketitle

\begin{abstract}
This work presents a new detailed study on the energy-dependent variation in the X-ray polarisation of the Crab Pulsar Wind Nebula (PWN), obtained using data from the Imaging X-ray Polarimetry Explorer (\textit{IXPE}).
For the entire PWN, we observed a linear variation in polarisation degree (PD), and detected the rotation of the polarisation angle (PA) with the energy at higher than 99.9999\% of the confidence level. 
This energy-dependent polarisation variation is in line with the indication found in Vela PWN by \textit{IXPE}, and it can be interpreted as the emitting region of the polarised photons shrinks with increasing energy, leading to higher PD because they are less influenced by the turbulence of the magnetic field.
We compared the \textit{IXPE} polarisation results with those of other hard X-ray/gamma observatories (\textit{PoGO+, Intregral, AstroSat}) for the PWN, finding the same trend from soft-X to hard-X with the PD increasing with the energy and the PA approaching the pulsar's spin axis. In fact, in this wide energy band, the fitting results show an energy trend for the PA compatible with the estimated pulsar's spin axis within 3$\sigma$ of confidence level.

\end{abstract}




\section{Introduction}

The Crab pulsar (PSR B0531+21) and its nebula are among the most fascinating and mysterious high-energy astrophysical objects in our universe. This pulsar was born from the cataclysmic explosion of a massive star, known as supernova SN1054, which occurred more than one thousand years ago \citep{1972VA.....13....1H,1995VA.....39..363B}.

The Crab is a complex astrophysical system that consists of several key components. \textit{Chandra} X-Ray Observatory has shown that, beyond the central pulsar, which serves as the primary energy source, there is a diffuse pulsar wind nebula (PWN) that extends outward and other distinct structures such as a jet and a torus located within the inner nebula \citep{2000ApJ...536L..81W}, observed also in Optical observations \citep{Hester+1995, Hester+2002}.
The Crab pulsar (PSR) emits its energy primarily through a relativistic flow of magnetised plasma \citep{doi:10.1126/science.304.5670.531}.
This pulsar wind is thought to be predominantly composed of electron-positron pairs. These pairs flow outward and interact with the remaining ejecta of the progenitor star. As these particles spread out in the nebula, they lose energy due to synchrotron and inverse Compton radiation, creating the glowing PWN observed today.
 
The Crab PWN is an ideal target for studying polarisation because of its exceptional brightness over the whole electromagnetic wavelength.
In the radio band, at 89.198\,GHz, its polarisation degree (PD) of the Crab nebula is $(8.8 \pm 0.2)\%$ and the polarisation angle (PA) is $(149.9 \pm 0.2)^\circ$ \citep{2010A&A...514A..70A}.
Moving to the optical band, polarimetric observations of the Crab PWN reported a PD of $\sim$70$\%$ with a PA of $\sim$118$^\circ$ in \cite{1981MNRAS.196..943J}, while a PD of $\sim$47$\%$ with a PA of $\sim$130$^\circ$ were measured by \cite{1988MNRAS.233..305S}.
In the X-ray band, Polarlight CubeSat in 2020 \citep{2020NatAs...4..511F} confirmed the classic results obtained in 1976 and 1978 by the OSO-8 mission \citep{1976ApJ...208L.125W,1978ApJ...220L.117W}, that the PD of the PWN after removing the contribution of the pulsar was around 19--20\%.
More recently, Imaging X-ray Polarimetry Explorer (\textit{IXPE}) observations of the Crab nebula improve the previous results in the soft X-ray band and indicate that the Crab nebula, including its pulsar, has an integrated spatial polarisation of $(20.2 \pm 0.36)\%$
with a PA of $(145.6 \pm 0.51)^\circ$. Moreover, \textit{IXPE} allows for the first spatially resolved polarisation map in the 2--8\,keV energy band. The \textit{IXPE} polarisation map suggests a predominantly toroidal magnetic field, with noticeable variations between the northern and southern regions of the nebula \citep{2023NatAs...7..602B}. 
The PD decreases by ${\sim}6\%$ along the southern jet, while the magnetic field orientation shifts from perpendicular to parallel relative to the jet axis \citep{2024ApJ...973..172W}.
Extending to the hard X-ray band, the polarisation observations provided by \textit{PoGo+} in the 20--160\,keV show that the PD of the PWN + PSR is $(20.9 \pm 5)\%$ and PA is $(131.3 \pm 6.6)^\circ$.
In the phase-resolved analysis of \textit{PoGo+}, the pulsar and nebula components are separated by isolating the off-pulse (OP) emission, which corresponds to the interval between the end of the interpulse and the start of the main pulse, when the pulsar flux reaches its minimum. The PD of OP is $(17.4 \,^{+8.6}_{-9.3})\%$, and the PA is $(137 \pm 15)^\circ$ \citep{Chauvin2017}. These results demonstrate that the Crab PWN exhibits a typical PD of ${\sim}20\%$ and PA of $\sim$145$^\circ$ in the high energy band.

Compared with previous X-ray polarisation detectors, the \textit{IXPE}'s excellent polarisation sensitivity allows for a more detailed study of the energy-resolved polarisation in the soft X-ray band for the Crab PWN. For example, this approach has been applied to Vela PWN. The \textit{IXPE} observation showed a slight increase in PD with energy that was attributed to the fact that the emission region contracts with increasing energy, thus photons arising from a smaller range of magnetic field orientations \citep{2022Natur.612..658X}. 
In this paper, we focus on a new detailed energy-resolved polarimetric analysis of Crab PWN, combining all the observations performed by \textit{IXPE} up to now. We organised the paper as follows. In Section 2, we describe the data used and how the polarimetric analysis is performed, identifying the Crab PWN region, and reporting the polarisation properties as a function of the energy. In Section 3, we present a discussion of our results and conclusions.

\section{Data reduction and results}
\subsection{\textit{IXPE} observations and data reduction}
On 9 December 2021, NASA and the Italian Space Agency (ASI) successfully launched \textit{IXPE} \citep{10.1117/1.JATIS.8.2.026002,2021AJ....162..208S}.
This mission marks the beginning of an unprecedented exploration of the X-ray universe, made possible by three identical X-ray telescopes equipped with polarisation-sensitive detector units operating in the 2--8\,keV energy range \citep{Costa+2001Natur.411..662C,Baldini+2021APh,2022AJ....164..103D}. \textit{IXPE} opens a new window into the high-energy cosmos, promising to reveal hidden celestial phenomena and provide groundbreaking insights.

\textit{IXPE} has observed the Crab over four different periods: (1) February 21–22, 2022, and March 7–8, 2022; (2) February 22–24, 2023, and April 1–3, 2023; (3) October 9–10, 2023; and (4) August 19–20, 2024. The total exposure time is $\sim$371\,ks. The observation IDs are 01001099, 02001099, 02006001 and 03009601, respectively. The calibrated data of those observations have been released through the \textsc{HEASARC}. We used HEASoft software\footnote{https://heasarc.gsfc.nasa.gov/lheasoft/} (version 6.31.1) \citep{heasoft} to analyse \textit{IXPE} data jointly with \textit{IXPE} collaboration software, named \textsc{ixpeobssim} V31.0.1\footnote{https://ixpeobssim.readthedocs.io/en/latest/} \citep{2022SoftX..1901194B}, with the response matrix updated to version 13, to analyse the polarisation properties of the Crab PWN. In particular, several preprocessing steps were performed to ensure the accuracy of the data, including energy calibration (only for the first observation, as required in the obsID prescription), detector World Coordinate System correction, and filtering out bad time intervals (such as those related to solar activities). The particle and instrumental background events were removed following the \cite{Alessandro+2023AJ....165..143D} algorithm. 
Barycentric corrections for \textit{IXPE} events were made using the \texttt{barycorr} tool in HEASoft v6.30.1, and the JPL-DE430 solar-system ephemeris was utilised. The Jodrell Bank Observatory ephemeris used for each observation epoch is detailed in Table \ref{table:ephemeride}.
The core of the main peak for the pulse profile of the Crab pulsar is located in the phase of 0.13 (see Appendix \ref{app:off-pulse}), the same as that used \cite{2023NatAs...7..602B}.
After all these processing steps, 
the polarisation (unweighted) analysis \citep{DiMarco_2022} was then performed using the \textsc{pcube} algorithm in \textsc{ixpeobssim}, which allowed us to obtain model-independent polarisation results across different energy bands and is based on \citet{Kislat15}.

To focus our polarimetric study on the Crab PWN and, therefore, to minimise the contamination of the pulsar, we select the PWN region and extracted data within the phase range of 0.7–1.0 for all subsequent analyses (see Fig. \ref{fig:offpulse}).
Since the Crab PWN is extremely bright (\text{\textit{IXPE} source counting rates} \textgreater \SI{2}{\text{cps} \,\text{arcmin}^{-2}}), the background is negligible and following the prescriptions reported in \cite{Alessandro+2023AJ....165..143D}, all polarimetric analyses were performed without background subtraction.

\subsection{Results}

\begin{figure}
    \centering
    \includegraphics[width=1.0\columnwidth]{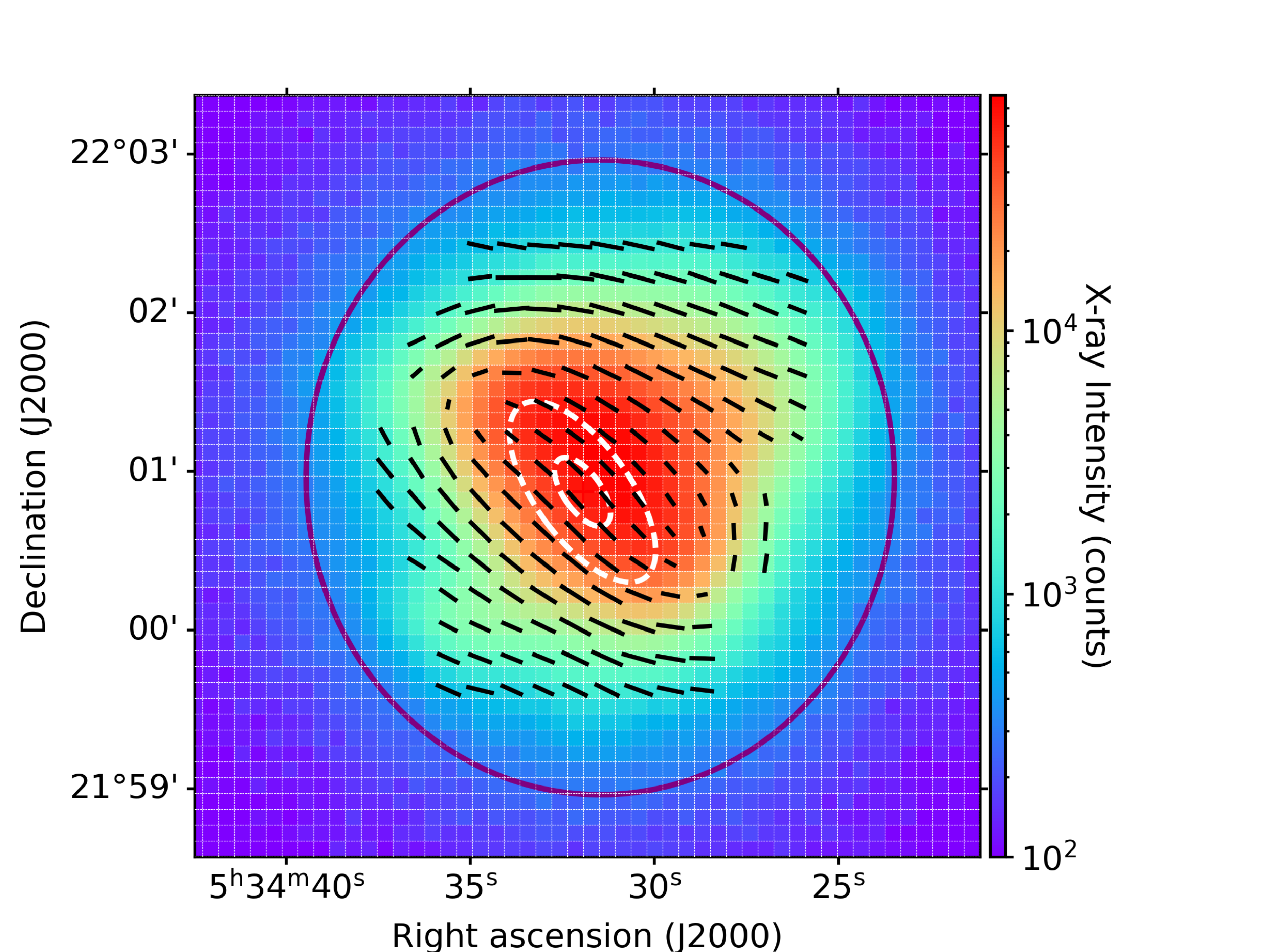}
    \caption{ 
    The intensity map of the Crab PWN using 
    the off-pulse photons in the 2--8\,keV energy band. The image is overlaid with the reconstructed
    direction of the magnetic field obtained considering the direction orthogonal to the measured PA. The purple circular region with a radius of $120''$ centred on the pulsar represents the PWN region in our study. The white ellipses represent the double-ring structure of the Crab PWN \citep{2004ApJ...601..479N}.
    }
    \label{fig:reg_select}
\end{figure} 

We obtained the polarisation maps using a top-hat kernel implemented in \textsc{ixpeobssim}.
We used a kernel of 5 $\times$ 5 pixels, each pixel representing a region of $6'' \times 6''$. This kernel \citep{Deng+2024}, therefore, covers a total region of $30'' \times 30''$, comparable to the angular resolution of \textit{IXPE} \citep{10.1117/1.JATIS.8.2.026002}.
Smoothed analysis was applied to each $6'' \times 6''$ pixel with a step length of 5 pixels, allowing the intensity maps of the Crab PWN to be obtained in the 2--8\,keV energy range, as shown in Fig.\ref{fig:reg_select}. This map is overlaid with the reconstructed polarisation vectors of the magnetic field with detection significance $>$5$\sigma$, showing a pattern compatible with a toroidal magnetic field similar to those found in \cite{2023NatAs...7..602B}.
The circular region, with a radius of $120''$ centred on the Crab pulsar, represents the PWN region in our study. 
Selecting the off-pulse phase (0.7-1.0), the whole PWN has PD of $(20.44 \pm 0.19)\%$ and PA of $(145.63 \pm 0.26)^\circ$ in 2--8\,keV.

We also performed the data analysis for the Crab PWN by integrating all the photons within the defined PWN region except for in the centre 20"-radius circle, which is the pulsar region. In this case, PD is calculated at $(19.49 \pm 0.11)\%$ and PA is $(148.91 \pm 0.16)^\circ$.
For the energy-integrated result, using the off-pulse photons, we get a slightly higher PD due to the fact that the contamination of the pulsar is removed. In fact, the pulsar could contaminate the external PWN region \citep{2023SPIE12679E..1CK}, and therefore, the best method to study the polarisation of the PWN is to use the off-pulse photons, for which the contribution of pulsar is negligible.

\begin{figure}
    \centering
    \includegraphics[width=1.0\columnwidth]{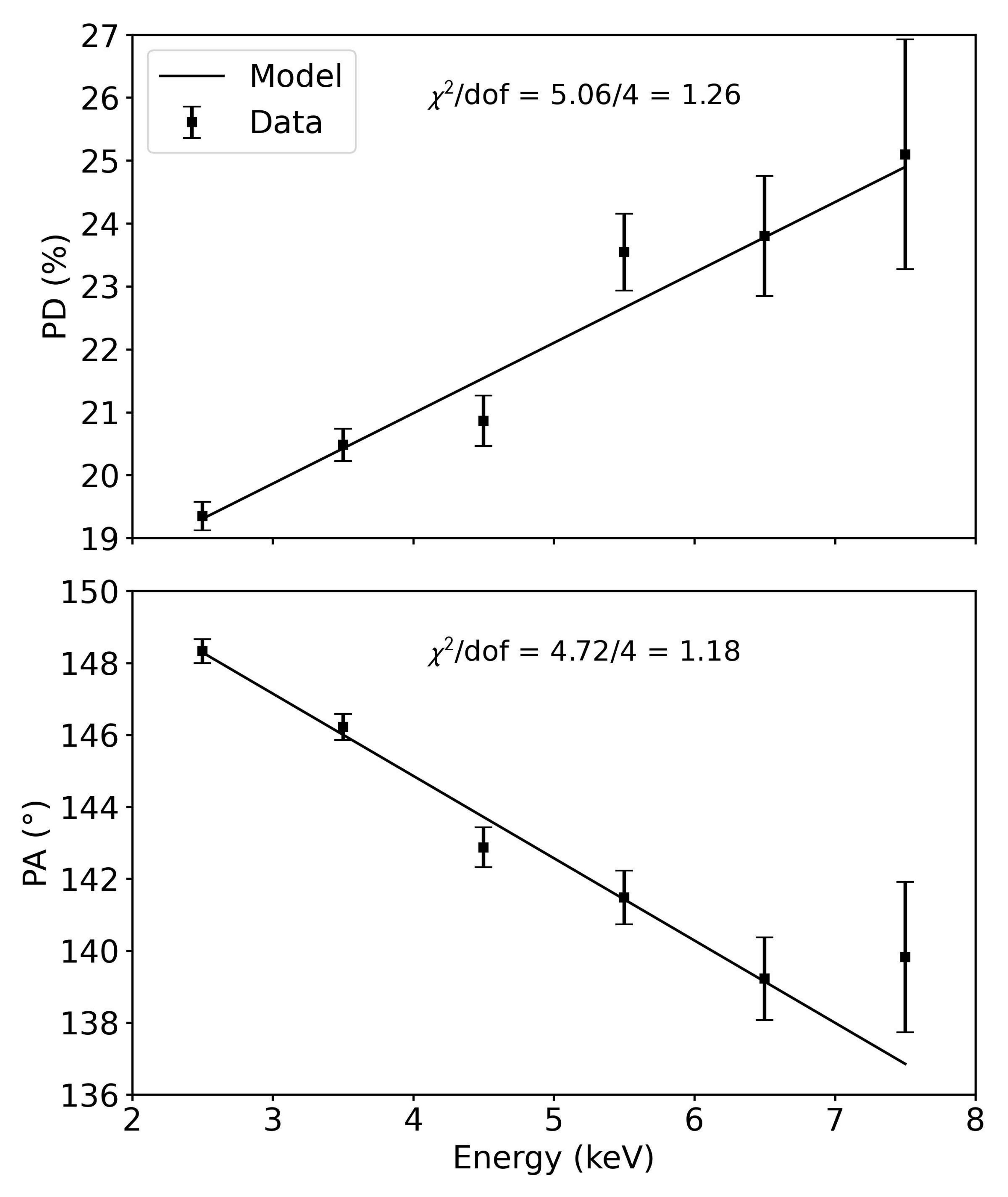}
    \caption{
    PD (top) and PA (bottom) of Crab PWN as a function of the energy. Each data point represents a 1\,keV bin-width. Results were obtained with uncertainties reported at 68$\%$ confidence level. Black line represents the linearly fitted model.}
    \label{fig:fit}
\end{figure}

\begin{table}
\normalsize
\centering
\caption{Polarisation results of the Crab PWN in different energy ranges. Errors are reported at 68\% confidence level.}
\begin{adjustbox}{max width=\textwidth}
\begin{tabular}{lcc}
\toprule
Energy (keV) & PD (\%) & PA ($^\circ$) \\
\midrule
\multicolumn{1}{c}{2--3} & 19.35 $\pm$ 0.23 & 148.33 $\pm$ 0.33 \\
\multicolumn{1}{c}{3--4} & 20.48 $\pm$ 0.26 & 146.21 $\pm$ 0.36 \\ 
\multicolumn{1}{c}{4--5} & 20.86 $\pm$ 0.40 & 142.87 $\pm$ 0.55 \\
\multicolumn{1}{c}{5--6} & 23.54 $\pm$ 0.61 & 141.48 $\pm$ 0.74 \\ 
\multicolumn{1}{c}{6--7} & 23.80 $\pm$ 0.95 & 139.22 $\pm$ 1.14 \\
\multicolumn{1}{c}{7--8} & 25.10 $\pm$ 1.82 & 139.82 $\pm$ 2.09 \\ 
\midrule
\multicolumn{1}{c}{3--5} & 20.58 $\pm$ 0.22 & 145.01 $\pm$ 0.31 \\
\multicolumn{1}{c}{5--8} & 23.95 $\pm$ 0.59 & 140.42 $\pm$ 0.71 \\
\multicolumn{1}{c}{2--8} & 20.44 $\pm$ 0.19 & 145.63 $\pm$ 0.26 \\ 
\bottomrule
\end{tabular}
\end{adjustbox}
\label{table:summary_pol_2-8}
\end{table}

\begin{figure}
    \centering
    \includegraphics[width=0.95\columnwidth]{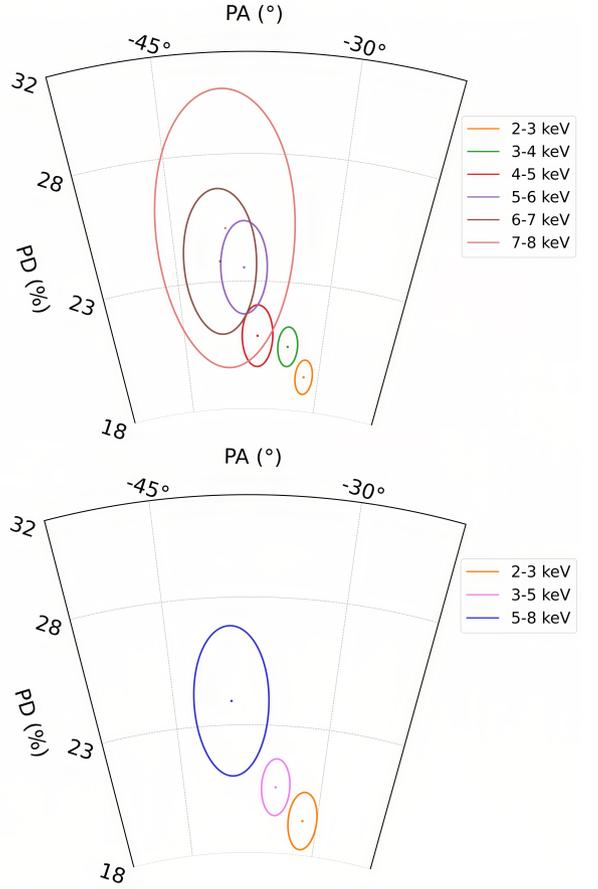}
    \caption{
     Top panel: the energy-resolved polarisation properties of Crab PWN in the PD-PA polar plane with uncertainties calculated at 99.7$\%$
     confidence level. The six energy bins are the same of the Fig. \ref{fig:fit}.
     Bottom panel: the same as above but with the uncertainties calculated at 99.9999$\%$ confidence level and the high energy bins are merged into 3--5\,keV and  5--8\,keV bin to increase their significance.
    }
    \label{fig:pwn_map2}
\end{figure}   

We studied the polarisation properties as a function of the energy.
The energy-resolved polarisation results of the PWN, 
are reported in Table \ref{table:summary_pol_2-8} and Fig. \ref{fig:fit}.
We divided the \textit{IXPE} energy band into 1\,keV energy bins, finding that the PD of PWN increases with energy: from $(19.35 \pm 0.23)\%$ in the 2--3 keV energy range to $(25.10 \pm 1.82)\%$ in the 7--8 keV energy range. Meanwhile, the PA rotates from $(148.33 \pm 0.33)^\circ$ to $(139.82 \pm 2.09)^\circ$. 
To verify this variation of polarisation with the energy, we first fit the PD and PA with a constant function. The reduced chi-squares are $\chi^2/\text{dof}=70.6/5=14.1$ for PD and $\chi^2/\text{dof}=168.5/5=33.7$ for PA, showing that both PD and PA are not consistent with a constant energy independent value.
Thus, we performed a linear fit on the data, with the results shown in Fig. \ref{fig:fit}. 
In this case, the reduced chi-square for PD is $\chi^2/\text{dof}=5.06/4=1.26$, and for PA it is $\chi^2/\text{dof}=4.72/4=1.18$.
The linear trend describes the PA and PD behaviours well with energy.

In order to better consider the error correlation between PD and PA \citep[see, e.g.,][]{DiMarco_2022}, we present the results in the PD-PA polar plane (see Fig. \ref{fig:pwn_map2}) which gives the correct confidence level for a polarimetric measurement, giving a more robust evaluation of the uncertainties.
In the upper panel, the contours are at the 99.7\% confidence level for each 1\,keV energy bin value. The last four bins are not statistically distinguishable from each other due to the larger errors.
We thus merge the last five bins (3--4\,keV, 4--5\,keV, 5--6\,keV, 6--7\,keV, 7--8\,keV) into two bins, obtaining a value with reasonable errors in the 3--5\,keV and 5--8\,keV energy range; the result is shown in the bottom panel of Fig. \ref{fig:pwn_map2}. 
Instead of contours at 99.7\% confidence level, we draw contours at 99.9999\% confidence level for the three energy bins, and their contours have no overlaps due to the improvement in significance. We observed a significant X-ray PA swing with the energy for the Crab PWN.

\begin{figure}
    \centering
    \includegraphics[width=0.55\textwidth]{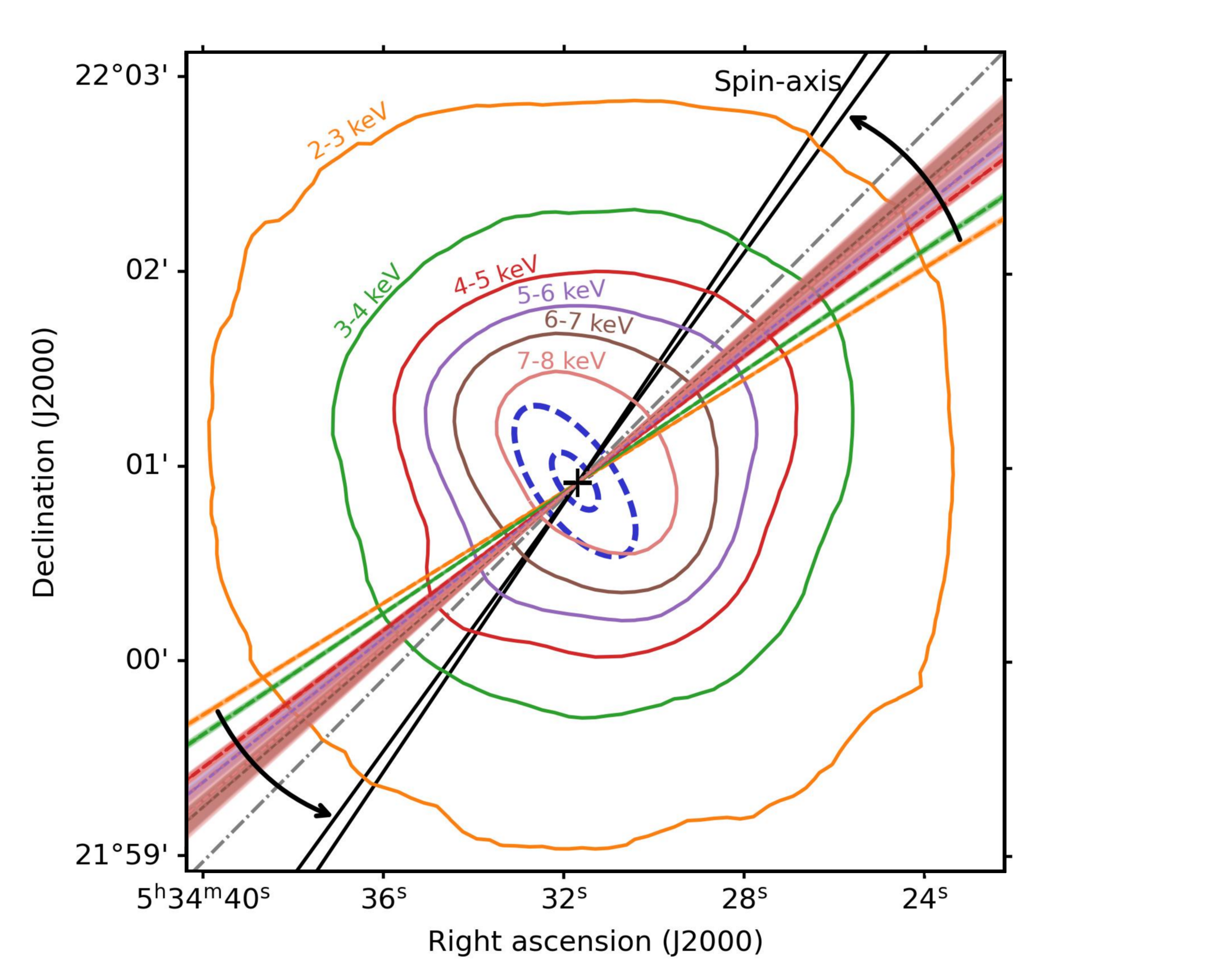}
  \caption{Image of the Crab PWN intensity contours obtained with \textit{IXPE} data in different energy bands (2--3, 3--4, 4--5, 5--6, 6--7, and 7--8\,keV) using off-pulse photons, and shown in different colours. The PAs and their errors calculated within these energy ranges, as listed in Table \ref{table:summary_pol_2-8}, are presented with the dash lines and shadow region in the same colours as the contours, noting that the PAs in the 6--7 and 7--8\,keV energy bands are too close to be separated.
  The blue ellipses represent the double-ring structure of the Crab PWN and the black solid line represents the pulsar's rotational axis \citep{2004ApJ...601..479N}, and the grey dash-dotted line represents the asymptotic angle derived from PA fitting shown in Fig. \ref{fig:pol_ene}. 
  }
  \label{fig:contour}
\end{figure}


\begin{figure*}
    \centering
    \includegraphics[width=0.75\textwidth]{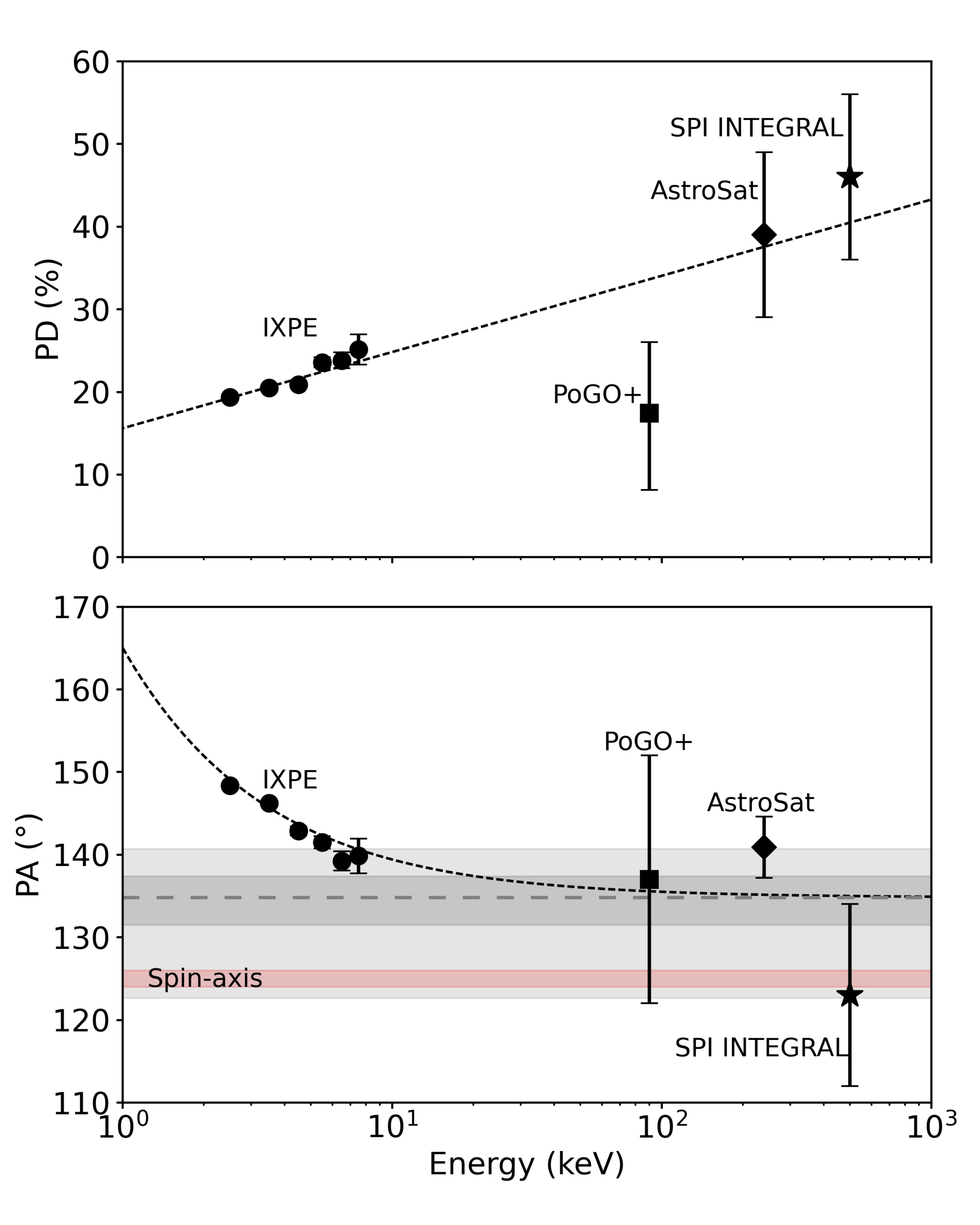}
    \caption{
    Polarisation properties of the Crab PWN in function of energy. 
    Top panel: The PD as a function of energy from soft to hard X-ray bands with different observatories. The black dashed line is the fitting model for the PD, represented by a linear function. 
    Bottom panel: The PA as a function of energy with different observatories.
    The black dashed line is the fitting model for the PA, represented by a power-law function.
    The grey loosely dashed line represents the asymptote obtained by the fit of the PA, which is \protect${\sim}135^\circ$ and the shaded gray region represents its error at 1$\sigma$ of confidence level. On top of this, we display the error region at 3$\sigma$ of confidence level in a lighter shade of gray. This value overlaps with the pulsar's rotation axis region, which is presented in shaded red as reported in \protect\cite{2004ApJ...601..479N}. 
    For \textit{IXPE} data, the energy-resolved polarisations in 1\,keV energy bin are reported as in Table~\protect\ref{table:summary_pol_2-8}. 
    For \textit{PoGO+}, the polarisation in the 20--160\,keV energy band is from \protect\cite{Chauvin2017}; 
    for \textit{AstroSat}, the polarisation in the 100--380\,keV energy band is from \protect\cite{2018NatAs...2...50V};
    anf for SPI Integral, the result in the 0.1--1\,MeV energy band is from \protect\cite{2008Sci...321.1183D}. All the polarimetric results for Crab PWN are calculated by using off-pulse photons.}
    
    \label{fig:pol_ene}
\end{figure*}

\section{Discussion and Conclusion} \label{sec:dis}

We analysed the data of all observations of the Crab PWN performed by \textit{IXPE} to study the energy-resolved polarisation for the first time, finding significant variation.

Within the 2--8\,keV energy range, for the whole PWN, we find that the PD increases and PA rotates with increasing energy, which are well described by a linear function.
\cite{2015ApJ...801...66M} reported the image of the intensity contours of the Crab PWN as a function of the energy in the X-ray band, concluding that the PWN emission gradually shrinks with increasing energy. 
We used \textit{IXPE} data to plot the PWN count map, and the intensity contours of the off-pulse photons at different energy ranges, as shown in Fig. \ref{fig:contour}. We applied a Stokes I threshold of 1530 counts, which is determined by considering the results across different energy ranges and is close to the maximum value for the 7--8 keV contour.
The overall morphological consistency of X-ray contours across multiple energy bands is maintained, and a notable size reduction is observed with increasing photon energy. This trend is similar to the result reported in \cite{2015ApJ...801...66M} using \textit{NuSTAR} observations.
Incorporating with our polarisation results, it suggests that the high-energy photons mainly originated from the inner nebula close to the pulsar, where they were less influenced by the turbulence, leading to higher PD with respect to the low-energy photon dominated by emission from the outer nebula. 
This is confirmed by spectral analysis of the Crab nebula. In fact, observations performed by \textit{Chandra} X-ray Observatory \citep{2004ApJ...609..186M} show that the photon index ranges from 1.65 in the inner PWN regions (including the inner ring and torus) to 2.15 or higher toward the outer edges of the nebula. This means that the spectrum is harder in the inner nebula, and a higher proportion of high-energy photons locate close to the pulsar, which is in agreement with the contours reported in Fig. \ref{fig:contour}.

We also find that within the 2--8\,keV energy range, the PA of the PWN changes significantly with the energy. The variation in PA with energy exceeds 99.9999\% confidence level, as shown in Fig. \ref{fig:pwn_map2}.
We notice that the spin-axis position angle of the Crab PSR is estimated approximately between $124^\circ$ and $126^\circ$ \citep{2004ApJ...601..479N}. The energy-resolved PA seems rotates towards the pulsar’s rotation axis.

From the prediction of the MHD simulation \citep{2005A&A...443..519B,2006A&A...453..621D}, the integrated PA should align with the symmetry axis of the toroidal magnetic field near the termination shock. The power distribution of the pulsar wind is minimal along the spin axis and maximal in the equatorial direction. As a result, the termination shock is located in the equatorial plane, and the symmetry axis is perpendicular to this plane. Consequently, the PA is aligned with the pulsar's spin axis \citep{2002MNRAS.329L..34L,2002MNRAS.336L..53B}.
For the Crab PWN, termination shock is believed to be close to the inner ring (the smaller ellipse shown in Fig. \ref{fig:reg_select}), which is also the region having a harder spectrum. Thus, we could expect a better alignment with the integrated PA and the rotation axis in a higher energy range. To test it, we collect all the polarisation measurements of Crab PWN from the soft to the hard X-ray bands.
In the soft 2--8\,keV energy range, we use the energy-resolved polarimetric results of \textit{IXPE} discussed previously and listed in Table \ref{table:summary_pol_2-8}.
In higher energy ranges, \cite{Chauvin2017} reported a PD of $(17.4 \,^{+8.6}_{-9.3})\%$ and a PA of $(137 \pm 15)^\circ$ for the PWN using \textit{PoGO+} in the energy range of 20--160\,keV. \textit{AstroSat} reported for PWN a PD of $(39.0 \pm 10.0)\%$ and a PA of $(140.9 \pm 3.7)^\circ$ in the energy range 100--380\,keV \citep{2018NatAs...2...50V} and \textit{SPI} onboard the \textit{Integral} reported the PD of $(46 \pm 10)\%$ and the PA of $(123 \pm 11)^\circ$ for PWN in 0.1--1\,MeV energy range \citep{2008Sci...321.1183D}.
All these polarisation results and the energy-resolved polarisation of \textit{IXPE} are reported in Fig. \ref{fig:pol_ene}.
The linear function describes the PD behavior, with $\chi^2/\text{dof} = 12.36/7 = 1.76$,
corresponding to a null hypothesis probability of 9\%), so that the PD increases with increasing energy not only in the \textit{IXPE} band (see Fig.~\ref{fig:fit}) but also from the soft X-rays to the hard X-ray bands.
On the other hand, a linear function does not fit the PA values.
Since we already noticed that the PA change rate in the high energy band is slower, we used a power-law function, \( y = kx^{-\alpha} + b \), to fit the PA across the wide energy bands.
The best-fit parameters are obtained from a Markov Chain Monte Carlo (MCMC) with \(n_{\text{walker}} = 32 \),  a burn-in of $2\times10^3$, and a chain length of $1\times10^4$; the reduced chi-squared is $\chi^2/\text{dof}=15.36/6=2.56$, corresponding to a null hypothesis probability of 2\%.
More details are shown in Appendix Fig. \ref{fig:corner}.
The fitting yields an asymptote that the PA gradually approaches $135^\circ$.
This value is compatible at 3$\sigma$ confidence level with the pulsar's rotation axis \citep{2004ApJ...601..479N}. Furthermore, it is the best estimate that we could obtain based on the observations that have already been made up to now in the different energy bands. 
We notice that the PA values in the hard energy ranges, measured by \textit{PoGo+} (PA error of $15^\circ$), \textit{AstroSat} (PA error of $3.7^\circ$), and \textit{SPI INTEGRAL} (PA error of $11^\circ$) exhibit larger uncertainties with respect to \textit{IXPE} results. Additionally, data from 10 keV to several tens of keV are absent, thus several functions could fit the current data.
In fact, we also tried a broken-linear function that broke at roughly 6--7\,keV hints by \textit{IXPE} observation, which was able to fit the data as well. However, an abrupt change in PA is not naturally explained. 
Therefore, comparing the \textit{IXPE} results with the ones obtained at higher energy, we observe a similar trend extending from the soft to the hard X-ray band, where the PD increases with the energy and the PA rotates towards the spin axis of the pulsar.

In summary, this work provides a comprehensive analysis of the energy-dependent polarisation properties of the Crab PWN, revealing significant trends in PD and PA in multiple energy bands. Future work could focus on extending this analysis to additional PWNe and refining models to explain the observed variations in PA and PD. 
Similar behaviours have been reported in Vela PWN using \textit{IXPE} observation \citep{2022Natur.612..658X}, PD of the whole PWN increasing with increasing energy.
We also tried to study the energy-resolved polarisation of other PWNe observed by \textit{IXPE}. However, for MSH 15-52 \citep{2023ApJ...957...23R}, in the entire ``hand'' region, PDs are too low to be detected in different energy bands.
This is associated with the magnetic field structure of MSH 15-52, where the magnetic field lines are aligned along the filamentary structure of the nebula. 
Future additional observations will increase the statistics and possibly reveal similar polarisation trends.
Moreover, for PSR B0540-69 \citep{2024ApJ...962...92X}, due to its long distance and relatively low brightness, it is difficult to detect the energy-resolved polarisation variations of the PWN with the \textit{IXPE} spectropolarimetric capabilities.
The upcoming launch of the enhanced X-ray Timing and Polarimetry (\textit{eXTP}) Mission \citep{2019SCPMA..6229502Z} heralds the advent of future detectors, promises significant advances. 
With their significantly larger effective area, broader observational energy band, and improved angular resolution, both are expected to provide high-quality X-ray polarisation data across different PWNe for a better understanding of their connection with the central pulsar.

\section*{Acknowledgements}
This work is supported by National Key R\&D Program of China (grant No. 2023YFE0117200), and National Natural Science Foundation of China (grant No. 12373041 and No. 12422306), and Bagui Scholars Program (XF). This work is also supported by the Guangxi Talent Program (“Highland of Innovation Talents”). FLM is supported by the Italian Space Agency (Agenzia Spaziale Italiana, ASI) through contract ASI-INAF-2022-19-HH.0, by the Istituto Nazionale di Astrofisica (INAF) in Italy, and partially supported by MAECI with grant CN24GR08 “GRBAXP: Guangxi-Rome Bilateral Agreement for X-ray Polarimetry in Astrophysics”.

\section*{Data Availability}
This research used data products provided by the \textit{IXPE} Team (MSFC, SSDC, INAF, and INFN) and distributed with additional software tools by the High-Energy Astrophysics Science Archive Research Center (HEASARC, \url{https://heasarc.gsfc.nasa.gov}), at NASA Goddard Space Flight Center (GSFC). The Imaging X-ray Polarimetry Explorer (\textit{IXPE}) is a joint US and Italian mission.



\bibliographystyle{mnras}




\newpage

\appendix                  

\section{The pulse profile of Crab PSR}\label{app:off-pulse}

\begin{table*}
\normalsize
\centering
\caption{
JBO Ephemerides used in the phase folding of the \textit{IXPE} Crab observations. Ephemeris 1 was applied to ObsID 01001099, Ephemeris 2 to ObsID 02001099, Ephemeris 3 to ObsID 02006001 and Ephemeris 4 to ObsID 03009601.}

\begin{adjustbox}{max width=\textwidth}
\begin{tabular}{lcccc}
\toprule
Parameters & Ephemeris1 & Ephemeris2 & Ephemeris3 & Ephemeris4 \\
\midrule
PEPOCH (MJD) & 59625 & 59990 & 60202 & 60508 \\ 
$\nu$ (Hz) & 29.5870753202 & 29.5754926637 & 29.5687705364 & 29.5590746932 \\
$\dot{\nu}$ (10$^{-10}$ Hz s$^{-1}$) & -3.67470 & -3.67089 & -3.66897 & -3.66578 \\ 
$\Ddot{\nu}$ (10$^{-21}$ Hz s$^{-2}$) & 8.04 & 5.24 & 11.6 & 2.56 \\ 
\bottomrule
\end{tabular}
\end{adjustbox}
\label{table:ephemeride}
\end{table*}

The Jodrell Bank Observatory ephemeris we used is from \url{https://www.jb.man.ac.uk/pulsar/crab.html}, specifically in Table \ref{table:ephemeride}.

\begin{figure*}
    \centering
    \includegraphics[width=0.9\textwidth]{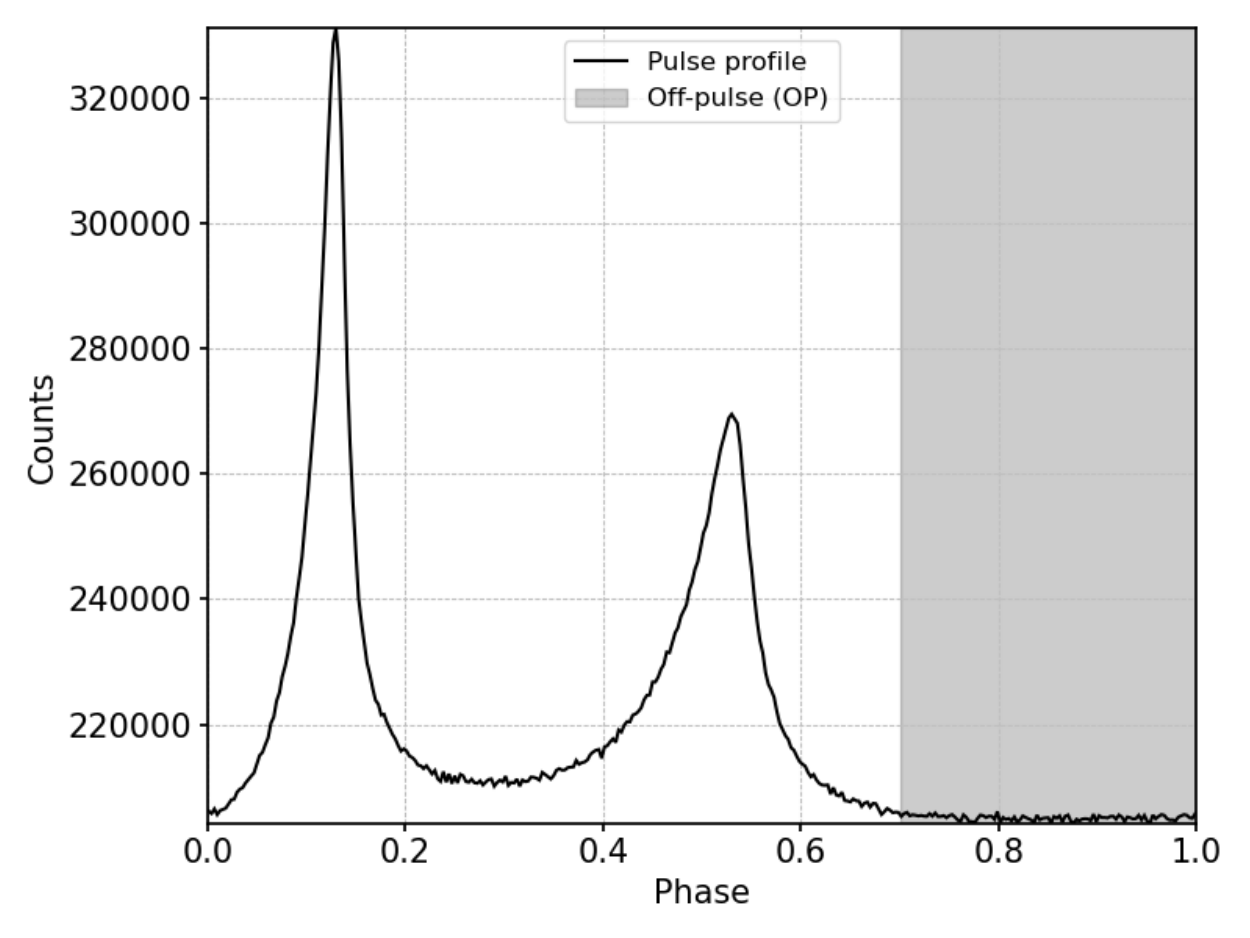}
  \caption{The pulse profile of the Crab pulsar with the main peak locates at 0.13, the gray shaded region represents the off-pulse phase selected in this work for the polarimetric analysis. 
  }
  \label{fig:offpulse}
\end{figure*}

Fig. \ref{fig:offpulse} depicts the X-ray pulse profile in 350 equally
spaced phase bins, the core of the main peak for the pulse profile of the Crab pulsar is located in the phase of 0.13. The so-called off-pulse region where we can consider the contribution of the pulsar negligible is defined in the phase interval 0.7--1.0.

\section{Best-fit parameters of PA as a function of the energy in the wide energy band}

\begin{figure*}
    \centering
    \includegraphics[width=0.9\textwidth]{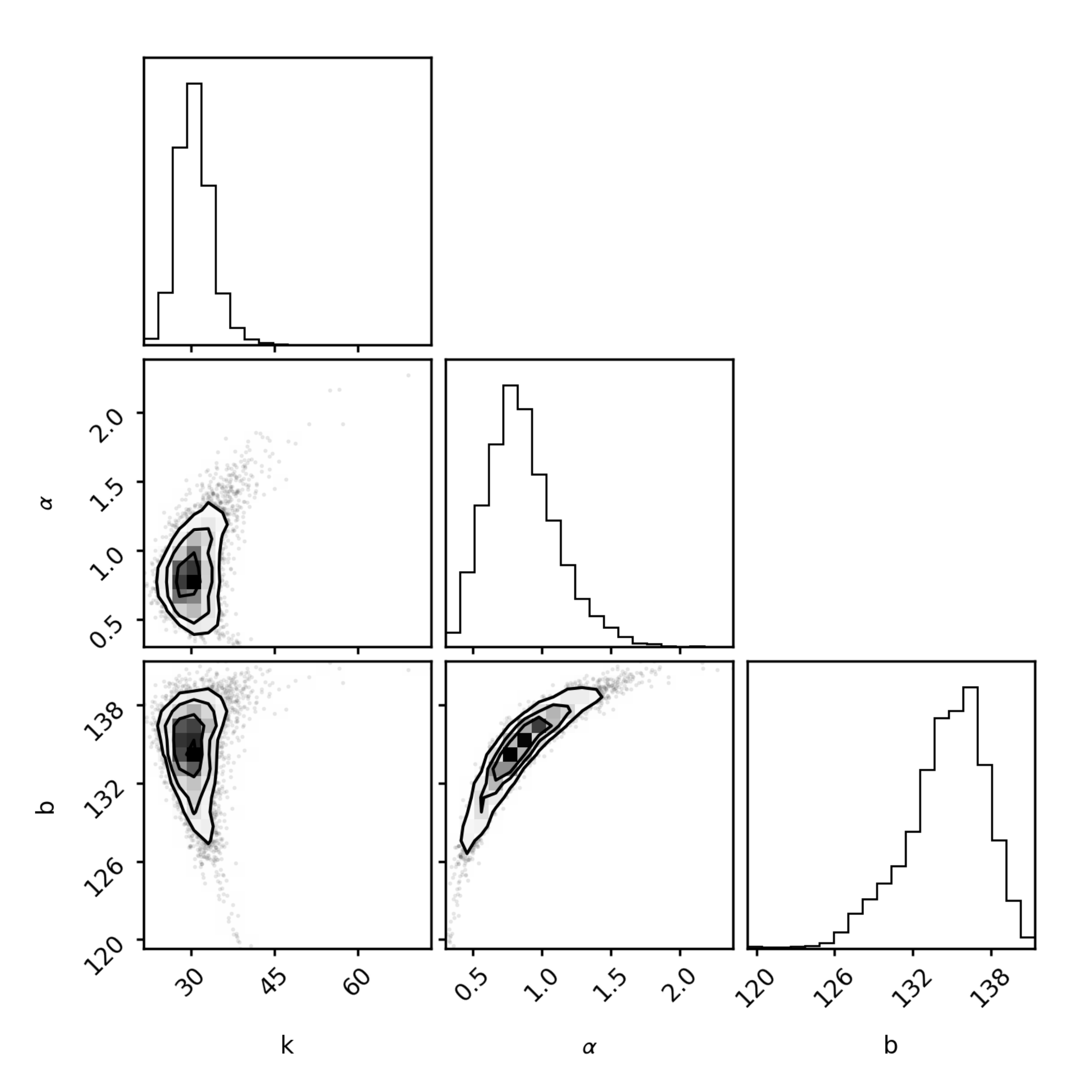}
  \caption{ Corner plots showing the optimal parameters for the PA fit reported in the bottom panel of Fig. \ref{fig:pol_ene}, with the fitted parameters at the 68\% confidence level \( k = (30.25^{+2.92}_{-2.70})^\circ\mbox{keV}^{-1} \), \( \alpha = 0.82^{+0.28}_{-0.20} \), and \( b = (134.79^{+2.60}_{-3.30})^\circ \).
  }
  \label{fig:corner}
\end{figure*}

Fig. \ref{fig:corner} shows the corner plots of the best-fit parameters of the power-law function, \( y = kx^{-\alpha} + b \), used to fit the PA from the soft to the hard X-ray energy band. It is worth noting that the b and $\alpha$ parameters are slightly correlated.

\bsp	
\label{lastpage}
\end{document}